\documentstyle[preprint,aps]{revtex}
\tightenlines
\RequirePackage{epsfig}
\def\Journal#1#2#3#4{{#1} {\bf #2}, #3 (#4)}
\def\PR{\em Phys. Rev.}
\def\PRL{\em Phys. Rev. Lett.}
\def\PRA{{\em Phys. Rev.} A}

\def\JMP{\em J. Math. Phys.}
\def\EPJD{{\em Eur. Phys. J.} D}
\def\RMP{\em Rev. Mod. Phys.}
\begin{document}
\draft
\title {Ground state properties of a one-dimensional condensate of
hard core bosons in a harmonic trap}
\author{M. D. Girardeau, E. M. Wright, and J.M. Triscari}
\address{Optical Sciences Center and Department of Physics\\
University of Arizona\\Tucson, AZ 85721}
\date{\today}
\maketitle
\begin{abstract}
The exact $N$-particle ground state wave function for a one-dimensional
condensate of hard core bosons in a harmonic trap is employed to obtain
accurate numerical results for the one-particle density matrix,
occupation number distribution of the natural orbitals, and momentum
distribution.  Our results show that the occupation of the lowest
orbital varies as $N^{0.59}$, in contrast to $N^{0.5}$ for a spatially
uniform system, and $N$ for a true BEC.
\end{abstract}
\vspace{0.2cm}

\pacs{03.75.Fi,03.75.-b,05.30.Jp}
\section{Introduction}
Recent advances in atom de Broglie waveguide technology
\cite{Key,Muller,Dekker,Thy,Hinds} and its potential applicability to
atom interferometry \cite{Berman} and integrated atom optics
\cite{Dekker,Sch} create a need for accurate theoretical modelling of
such systems in the low temperature, tight waveguide regime where
transverse excitations are frozen out and the
quantum dynamics becomes essentially one-dimensional (1D).
It has been shown by Olshanii \cite{Olshanii}, and also recently
by Petrov et al. \cite{PetShlWal00}, that at sufficiently low
temperatures, densities and large positive scattering length,
a Bose-Einstein condensate (BEC) in a
thin cigar-shaped trap has dynamics which approach those of a 1D
gas of hard core, or impenetrable, point bosons. This is a model for
which the exact many-body energy eigensolutions were found in 1960 using
an exact mapping from the Hilbert space of energy eigenstates of an
{\em ideal} gas of fictitious spinless fermions to that of many-body
eigenstates of hard core, and therefore {\em strongly interacting}, bosons
\cite{map,map2}. In this limit there are strong short-range pair
correlations which are omitted in the Gross-Pitaevskii (GP) approximation.
In the absence of a trap potential it is known \cite{Lenard} that the
occupation of the lowest orbital is of order $\sqrt{N}$ where $N$ is the
total number of atoms, in contrast to $N$ for the ideal Bose gas and GP
approximation.  Nevertheless, this system exhibits some BEC-like behavior
such as Talbot recurrences following an optical lattice pulse \cite{Rojo}
and dark soliton-like behavior in response to a
phase-imprinting pulse \cite{soliton}.

The case of harmonically trapped, hard core bosons in 1D is more relevant
to recent atom waveguide experiments \cite{BonBurDet00}. The spatial density
profile of the single-particle density is expressible in closed form, and
has recently been shown \cite{Kolomeisky} to be well approximated by
a modified 1D effective field theory, although we have recently shown
in a numerically accurate time-dependent calculation \cite{breakdown} 
that spatial interference of separated and recombined condensates
is much weaker than that predicted by the corresponding time-dependent
mean field theory \cite{Kolomeisky}.
Although the Fermi-Bose mapping theorem \cite{map,map2} implies that all
physical properties expressible in terms of spatial configurational
probabilities are the same for the actual bosonic system and the
fictitious "spinless fermion" system used for the mapping, the momentum
distribution of the bosonic system, or more generally its
occupation distribution over the relevant orbitals for a given geometry,
is very different in the bosonic system. It is known
\cite{Olshanii,Lenard,Vaidya} that for a spatially uniform
system of hard core bosons in 1D, the momentum distribution is
strongly peaked in the neighborhood of zero momentum, whereas that of the
corresponding Fermi system is merely a filled Fermi sea. In the case of
hard core bosons in a 1D harmonic trap, it is an interesting and
previously unanswered question whether the system undergoes true BEC or
merely an attenuated one such as that in the uniform system.  Ketterle
and Van Druten \cite{KetVan96} have shown that true BEC occurs for a
finite number of atoms in a 1D harmonic oscillator (HO) for an ideal gas.
We examine the question for hard core bosons in a 1D HO
by using the Fermi-Bose mapping theorem to
generate the exact many-body ground state. The most fundamental definition
of BEC and the condensate orbital is based on the large distance behavior
of the one-particle reduced density matrix $\rho_{1}(x,x')$. If off-diagonal
long-range order (ODLRO) is present and hence the largest eigenvalue of
$\rho_1$ is macroscopic (proportional to $N$) then the system is said to
exhibit true BEC and the corresponding eigenfunction, the condensate
orbital, plays the role of an order parameter \cite{PO,Yang2}. Although the
precise definition of ODLRO requires a thermodynamic limit not strictly
applicable to mesoscopic traps, the GP approximation assumes from the start
that ODLRO and macroscopic occupation of a single orbital are good
approximations in a trap, so examination of this assumption in the
Olshanii limit \cite{Olshanii} is important. In the remainder of this
paper we shall determine the many-body ground state and its salient
features, including the one-particle reduced density matrix and its
eigenvalues (occupation number distribution function) and eigenfunctions
(natural orbitals), as well as the momentum distribution function.
\section{Exact ground state wave function}
The Hamiltonian of N bosons in a 1D harmonic trap is
\begin{equation}\label{eq1}
\hat{H}=\sum_{j=1}^{N}
\left[-\frac{\hbar{^2}}{2m}\frac{\partial^2}{\partial x_{j}^{2}}
+\frac{1}{2}m\omega^{2}x_{j}^{2}\right]  .
\end{equation}
We assume that the two-body interaction potential consists only of a hard
core of 1D diameter $a$. This is conveniently treated as a constraint on
allowed wave functions $\psi(x_{1},\cdots,x_{N})$ such that
\begin{equation}\label{eq2}
\psi=0\quad\text{if}\quad |x_{j}-x_{k}|<a\quad,\quad 1\le j<k\le N  ,
\end{equation}
rather than as an infinite interaction potential. It follows from the
Fermi-Bose mapping theorem \cite{map,map2,soliton}
that the exact N-boson ground state $\psi_{B0}$ of the Hamiltonian (1)
with the constraint (2) is
\begin{equation}\label{eq3}
\psi_{B0}(x_{1},\cdots,x_{N})=|\psi_{F0}(x_{1},\cdots,x_{N})|  ,
\end{equation}
where $\psi_{F0}$ is the ground state of a fictitious system of $N$
spinless fermions with the same Hamiltonian (\ref{eq1}) and constraint.
At low densities it is sufficient \cite{Olshanii,PetShlWal00} to
consider the case of impenetrable point particles, the zero-range limit
$a\rightarrow 0$ of (2). Since wave functions of "spinless fermions"
are antisymmetric under coordinate exchanges,
their wave functions vanish automatically whenever any $x_{j}=x_{k}$,
the constraint has no effect, and the corresponding fermionic ground state
is the ground state of the {\em ideal} gas of fermions, a Slater determinant
of the lowest $N$ single-particle eigenfunctions $\phi_n$ of the harmonic
oscillator (HO)
\begin{equation}\label{eq4}
\psi_{F0}(x_{1},\cdots,x_{N})=\frac{1}{\sqrt{N!}}
\det_{(n,j)=(0,1)}^{(N-1,N)}\phi_{n}(x_{j}) .
\end{equation}
The HO orbitals are
\begin{equation}
\varphi_{n}(x)= \frac{1}
{\pi^{1/4}x_{osc}^{1/2}\sqrt{2^{n}n!}}e^{-Q^{2}/2}H_{n}(Q)
\end{equation}
with $H_n(Q)$ the Hermite polynomials and
$Q=x/x_{osc},x_{osc}=\sqrt{\hbar/m\omega}$
being the ground state width of the harmonic trap for a single atom.
By factoring the Gaussians out of the determinant and carrying out
elementary row and column operations, one can cancel all terms in each
$H_n$ except the one of highest degree, with the result \cite{Aitken}
\begin{eqnarray}
\det_{(n,j)=(0,1)}^{(N-1,N)}H_{n}(x_{j})
& = & 2^{N(N-1)/2}\det_{(n,j)=(0,1)}^{(N-1,N)}(x_{j})^{n} \nonumber\\
& = & 2^{N(N-1)/2}\prod_{1\le j<k\le N}(x_{k}-x_{j})
\end{eqnarray}
Substitution into (3) then yields a simple but exact analytical
expression
of Bijl-Jastrow pair product form for the $N$-boson ground state:
\begin{equation}
\psi_{B0}(x_{1},\cdots,x_{N})=C_{N}\left[\prod_{i=1}^{N}e^{-Q_{i}^{2}/2}
\right]
\prod_{1\le j<k\le N}|x_{k}-x_{j}|
\end{equation}
with $Q_{i}=x_{i}/x_{osc}$ and normalization constant
\begin{equation}
C_{N}=2^{N(N-1)/4}\left (\frac{1}{x_{osc}} \right )^{N/2}
\left[N!\prod_{n=0}^{N-1}n!\sqrt{\pi}\right]^{-1/2}  .
\end{equation}
It is interesting to
note the strong similarity between this exact 1D $N$-boson wave
function and the famous Laughlin variational wave function of the 2D ground
state for the quantized fractional Hall effect \cite{Laughlin},
as well as the closely-related wave functions for bosons with weak repulsive 
delta-function interactions in a harmonic trap in 2D found recently
by Smith and Wilkin \cite{SW}.
\section{Ground state properties}
In this section we numerically evaluate the ground state properties of a
1D condensate of $N$ hard core bosons in a harmonic trap
using the exact many-body wave function of the previous section.
\subsection{Single particle density and pair distribution function}
Both the single particle density and pair distribution function depend
only on the absolute square of the many-body wave function, and since
$|\psi_{B0}|^{2}=|\psi_{F0}|^{2}$ they reduce to standard ideal Fermi
gas expressions. The single particle density, normalized to $N$, is
\begin{eqnarray}
\rho(x) &=& N\int |\psi_{B0}(x,x_{2},\cdots,x_{N})|^{2}dx_{2}\cdots
dx_{N}\nonumber\\
& = & \sum_{n=0}^{N-1}|\varphi_{n}(x)|^{2}
\end{eqnarray}
We shall not exhibit it here since it has recently been calculated by
Kolomeisky {\it et al.} \cite{Kolomeisky}; see also our recent discussion of
the time-dependent case \cite{breakdown}.
The pair distribution function, normalized to $N(N-1)$, is
\begin{eqnarray}
& & D(x_{1},x_{2})=N(N-1)\int
|\psi_{B0}(x_{1},\cdots,x_{N})|^{2}dx_{3}
\cdots dx_{N} \nonumber\\
& & =\sum_{0\le n<n'\le N-1}\hspace{-0.5cm}
|\varphi_{n}(x_{1})\varphi_{n'}(x_{2})
-\varphi_{n}(x_{2})\varphi_{n'}(x_{1})|^{2}
\end{eqnarray}
Physically, the pair distribution function is the joint probability
density that if one atom is measured at $x_1$ then a second measurement
immediately following the first finds an atom at $x_2$.
Noting that terms with $n=n'$, which vanish by antisymmetry, can be
formally added to the summation (9), one can rewrite the pair distribution
function in terms of the single particle density and a correlation function
$\Delta$:
\begin{eqnarray}
D(x_{1},x_{2}) & = & \rho(x_{1})\rho(x_{2})
-|\Delta(x_{1},x_{2})|^{2} \nonumber\\
\Delta(x_{1},x_{2}) & = &
\sum_{n=0}^{N-1}\varphi_{n}^{*}(x_{1})\varphi_{n}(x_{2})
\end{eqnarray}
Although the Hermite polynomials have disappeared from
the expression (7) for the many-body wave function, they reappear upon
integrating $|\psi_{B0}|^{2}$ over $(N-1)$ coordinates to get the single
particle density $\rho(x)$ and over $(N-2)$ to get the pair distribution
function $D(x_{1},x_{2})$, and the expressions in terms of the HO orbitals
$\varphi_n$ are the most convenient for evaluation.

Figure \ref{Fig:one}
shows a gray scale plot of the dimensionless pair distribution function
$x_{osc}^2\cdot D(Q_1,Q_2)$ versus the normalized coordinates
$Q_{1,2}=x_{1,2}/x_{osc}$
for a) $N=2$, b) $N=6$, and c) $N=10$.  Some qualitative features
of the pair distribution function are apparent: In the first place it
follows either from the original expression (9) or from Eqs. (8) and (10)
that $D(x_1,x_2)$ vanishes at contact $x_1=x_2$, as it must because of
impenetrability of the particles, and we see this to be true in Fig.
\ref{Fig:one}.  Furthermore, the correlation term
$\Delta(x_{1},x_{2})$ is a truncated closure sum and approaches the
Dirac delta function $\delta(x_{1}-x_{2})$ as $N\rightarrow\infty$, as is to
be expected since the healing length in a spatially uniform 1D hard core
Bose gas varies inversely with particle number \cite{soliton}.
As a result the width of the null around the diagonal $Q_1=Q_2$
decreases with increasing $N$, and vanishes in the limit.  Away from
the diagonal along $Q_2=-Q_1$ the pair distribution function rises,
exhibits modulations for $N>2$, due to the oscillatory nature of the
HO orbitals, before decreasing back to zero at large distances.
For $|x_{1}-x_{2}|$ much larger than
the healing length, $D$ reduces to the uncorrelated density product
$\rho(x_{1})\rho(x_{2})$, so the spatial extent of the pair distribution
function is that of the density and varies as $N^{1/2}$ \cite{Kolomeisky}.
\subsection{Reduced single-particle density matrix}
The reduced single-particle density matrix with
normalization $\int\rho_{1}(x,x)dx=N$ is given by
%
\begin{eqnarray}
& & \rho_{1}(x,x')=N\int\psi_{B0}(x,x_{2},\cdots,x_{N}) \nonumber\\
& & \times \psi_{B0}(x',x_{2},\cdots,x_{N})dx_{2}\cdots dx_{N}
\nonumber\\
& & ={\cal N}_{N}e^{-Q^{2}/2}e^{-(Q')^{2}/2}
\int\prod_{i=2}^{N}e^{-Q_{i}^{2}}|Q_{i}-Q||Q_{i}-Q'| \nonumber\\
& & \times[\prod_{2\le j<k\le N}(Q_{k}-Q_{j})^{2}]dQ_{2}\cdots dQ_{N}  ,
\label{rho1}
\end{eqnarray}
with
%
\begin{equation}
{\cal N}_{N}=N\cdot 2^{N(N-1)/2}x_{osc}^{-1}
\left[N!\prod_{n=0}^{N-1}n!\sqrt{\pi}\right]^{-1}  .
\end{equation}
Although the multi-dimensional integral (\ref{rho1}) cannot be evaluated
analytically, it can be evaluated numerically by Monte Carlo integration
for not too large values of $N$ (the computing time scales as $N^4$).
Figure \ref{Fig:two}
shows a gray scale plot of the dimensionless reduced single-particle
density matrix
$x_{osc}\cdot\rho_1(Q,Q')$ versus the normalized coordinates
$Q$ and $Q'$ for a) $N=2$, b) $N=6$, and c) $N=10$.  We verified that
along the diagonal $\rho_1(Q,Q'=Q)=\rho(Q)$ reproduced the single-particle
density \cite{Kolomeisky}.  The off-diagonal elements of the
reduced density matrix relate to ODLRO, and it is clear that as
$N$ increases the off-diagonal elements are decreasing in contrast
to the diagonal.  This is a first indication that ODLRO vanishes
for a system of hard core bosons in a 1D HO in the thermodynamic
limit.
\subsection{ODLRO, natural orbitals and their occupation}
In a macroscopic system, the presence or absence of BEC is determined by the
behavior of $\rho_{1}(x,x')$ as $|x-x'|\rightarrow\infty$. Off-diagonal
long-range order is present if the largest eigenvalue of
$\rho_1$ is macroscopic (proportional to $N$), in which case the system 
exhibits BEC and the corresponding eigenfunction, the condensate
orbital, plays the role of an order parameter \cite{PO,Yang2}. Although
this criterion is not strictly applicable to mesoscopic systems, if the
largest eigenvalue of $\rho_1$ is much larger than one
then it is reasonable to expect that
the system will exhibit some BEC-like coherence effects. Thus we examine here 
the spectrum of eigenvalues $\lambda_j$ and associated eigenfunctions
$\phi_{j}(x)$ (``natural orbitals'') of $\rho_1$. Although natural orbitals
are a much-used tool in theoretical chemistry, they have only
recently been applied to mesoscopic atomic condensates \cite{DuBGly00}.
The relevant eigensystem equation is
%
\begin{equation}
\int_{-\infty}^{\infty}\rho_{1}(x,x')\phi_{j}(x')dx'=\lambda_{j}\phi_{j}(x)
\end{equation}
$\lambda_j$ represents the occupation of the orbital $\phi_j$, and one has
$\sum_{j}\lambda_{j}=N$. Numerical evaluation of the integral (13) by
discretization yields a readily-solved matrix eigensystem equation
which yields accurate numerical results for the largest eigenvalues
and associated eigenvectors. In Fig. \ref{Fig:three}(a) we show a
log-log plot of the fractional occupation of the lowest orbital
$f_0=\lambda_0/N$ versus the total particle number $N$ (solid line), along
with a best fit power-law $f_0\approx N^{-0.41}$ (dashed line).
This is to be contrasted
with the case of a spatially uniform system of hard core bosons for which
$f_0\approx N^{-0.5}$ \cite{Lenard}.  In both cases the fractional
occupation decreases with increasing $N$, and thus do not correspond to
true condensates for which $f_0=1$.  Nevertheless, the occupation of the
lowest orbital may still be large $\lambda_0\approx N^{0.59}$, and is
larger than the spatially uniform case $\lambda_0\approx N^{0.5}$, so
macroscopic quantum coherence effects reminiscent of BEC can still result
\cite{Olshanii,Lenard,Rojo,soliton,Kolomeisky,breakdown}.

Figure \ref{Fig:three}(b) shows the distribution of occupations $\lambda_j$
versus orbital number $j$ (the orbitals are ordered according to
eigenvalue magnitude, the largest eigenvalue being $j=0$) for
$N=2$ (circles), $N=6$ (stars), and $N=10$ (squares).  This figure
shows that as the lowest orbital occupation $\lambda_0$ increases with
increasing $N$ so does the range of significantly occupied higher-order
orbitals with $j>0$.  This means that the dominance of the lowest orbital
decreases with increasing $N$, so singling out $\phi_0(x)$ as a
macroscopic wave function for the whole system becomes more
problematic with increasing $N$ \cite{breakdown,Kolomeisky}.

The numerically computed lowest orbitals $\phi_0(Q)$ are shown in
Fig \ref{Fig:four} for a) $N=2$, b) $N=6$, and c) $N=10$, and they
show the expected broadening due to many-body repulsion as $N$ increases.
We remark that these lowest orbitals are not simply the square root of
the corresponding single-particle densities $\rho(Q)$ \cite{Kolomeisky}
as would be the case for a true condensate.
Figure \ref{Fig:five} shows the higher order orbitals $\phi_j(Q)$ versus
$Q$ for $j=1,2,3$ and $N=10$.  Although the higher-order orbitals
differ in detail from the HO orbitals they share the features that the
orbitals can be chosen real by removal of an overall phase, and that
the $j^{th}$ orbital has $j$ zeros.
\section{Momentum distribution}
For a spatially uniform system (no trap) the
natural orbitals are plane waves, so the occupation distribution of the
natural orbitals is the same as the momentum distribution. Although this
is not the case here due to the effect of the harmonic trap potential,
the momentum distribution is still physically important, so we study it here.
In terms of the boson annihilation and creation operators in
position representation (quantized Bose field operators) the one-particle
reduced density matrix is 
\begin{equation}
\rho_{1}(x,x')=\langle\Psi_{B0}|\hat{\psi}^{\dagger}(x')\hat{\psi}(x)|
\Psi_{B0}\rangle
\end{equation}
The momentum distribution function $n(k)$, normalized to 
$\int_{-\infty}^{\infty} n(k)dk=N$ , is 
$n(k)=\langle\Psi_{B0}|\hat{a}^{\dagger}(k)\hat{a}(k)|\Psi_{B0}\rangle$
where $\hat{a}(k)$ is the annihilation operator for a boson with momentum 
$\hbar k$. Then 
\begin{equation}
n(k)=(2\pi)^{-1}\int_{-\infty}^{\infty}dx\int_{-\infty}^{\infty}dx'
\rho_{1}(x,x')e^{-ik(x-x')}
\end{equation}
The spectral representation of the density matrix then leads to 
$n(k)=\sum_{j}\lambda_{j}|\mu_{j}(k)|^2$ where the $\mu_j$ are
Fourier transforms of the natural orbitals: 
$\mu_{j}(k)=(2\pi)^{-1/2}\int_{-\infty}^{\infty}\phi_{n}(x)e^{-ikx}dx$ .
Figure \ref{Fig:six} shows the numerically calculated dimensionless
momentum spectrum $k_{osc}\cdot n(\kappa)$ versus normalized
momentum $\kappa=k/k_{osc}$,
with $k_{osc}=2\pi/x_{osc}$, for  a) $N=2$, b) $N=6$, and c) $N=10$.
The key features are that the momentum spectrum maintains the
sharp peaked structure reminiscent of the spatially uniform
case \cite{Olshanii,Lenard} for the 1D HO, and that the peak becomes
sharper with increasing atom number $N$.  This is to be expected since
as the number of atoms increase the many-body repulsion causes the
system to become more spatially uniform within the trap interior.
\section{Summary and conclusions}
In summary, we have investigated the ground state properties of
a system of hard core bosons in a 1D HO using the exact many-body
wave function obtained using the Fermi-Bose mapping theorem.
Specifically, we have numerically evaluated the reduced single-particle
density matrix for the system using Monte-Carlo integration for
particle numbers up to $N=10$, and extracted several quantities of
physical significance, including the natural orbitals and momentum
spectrum.  Our main finding is that the lowest orbital occupation
scales as $\lambda_0\approx N^{0.59}$, so that the system does not
exhibit true BEC, counter to the case of an ideal gas in a 1D HO
\cite{KetVan96}.  Furthermore, this makes the introduction of an
order-parameter or macroscopic wave function for the whole system
more problematic for large $N$.  We have started to seek analytic
approaches to derive the observed scaling of the lowest orbital with
particle with no success so far.  We hope that these numerical results may
motivate others to approach this challenging problem.
\vspace{0.2cm}

\noindent
This work was supported by Office of Naval Research grant
N00014-99-1-0806.
\begin{figure}
\caption{Gray-scale plots of the dimensionless pair distribution function
$x_{osc}^2\cdot D(Q_1,Q_2)$ as a function of the dimensionless
coordinates $Q_1$ and $Q_2$, for a) $N=2$, b) $N=6$, and c) $N=10$.}
\label{Fig:one}
\end{figure}
\begin{figure}
\caption{Gray-scale plots of the dimensionless reduced
density matrix $x_{osc}\cdot\rho_1(Q,Q')$ as a function of the
dimensionless
coordinates $Q$ and $Q'$, for a) $N=2$, b) $N=6$, and c) $N=10$.}
\label{Fig:two}
\end{figure}
\begin{figure}
\caption{Occupation of the natural orbitals: a) fraction of atoms
in the lowest orbital $f_0=\lambda_0/N$ versus $N$, and b) 
$\lambda_j$ versus orbital number $j$ for
$N=2$ (circles), $N=6$ (stars), and $N=10$ (squares).}
\label{Fig:three}
\end{figure}
\begin{figure}
\caption{Lowest natural orbitals $\phi_0(Q)$ versus normalized
coordinate $Q$ for $a) N=2$, b) $N=6$, and c) $N=10$.}
\label{Fig:four}
\end{figure}
\begin{figure}
\caption{Higher-order natural orbits $\phi_j(Q)$ versus normalized
coordinate $Q$ for $N=10$ and a) $j=1$, b) $j=2$, and c) $j=3$.}
\label{Fig:five}
\end{figure}
\begin{figure}
\caption{Dimensionless momentum distribution $k_{osc}\cdot n(\kappa)$ versus
normalized momentum $\kappa=k/k_{osc}$ for $a) N=2$, b) $N=6$, and
c) $N=10$.}
\label{Fig:six}
\end{figure}
\newpage
\includegraphics*[width=0.4\columnwidth]{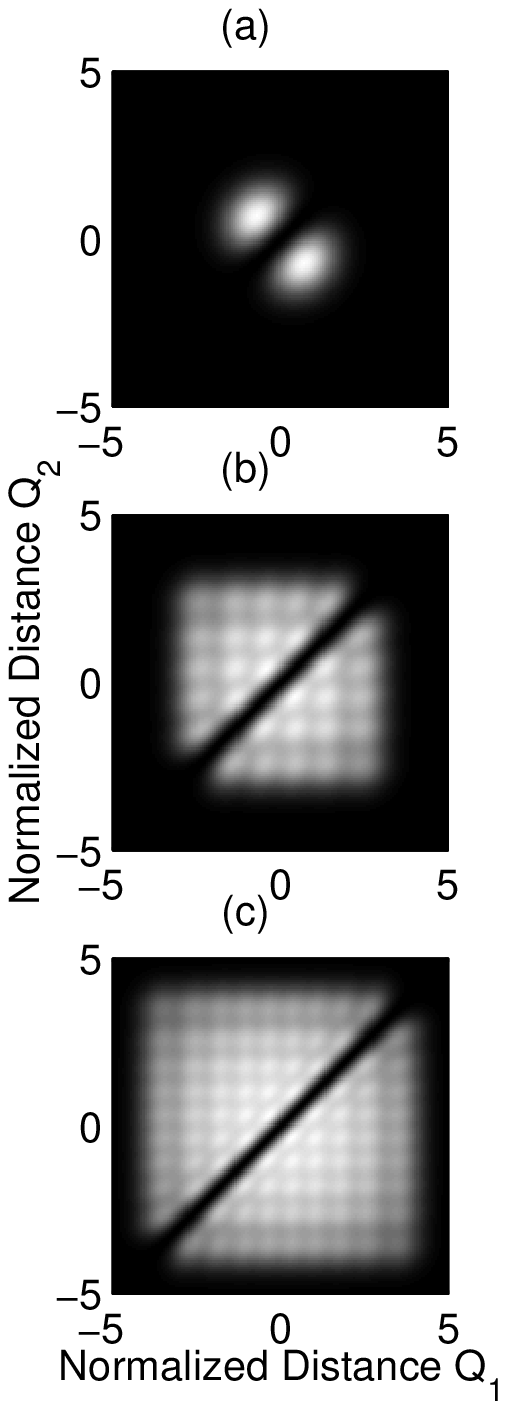}
\begin{center}
{\bf Figure 1}
\end{center}
\newpage
\includegraphics*[width=0.4\columnwidth]{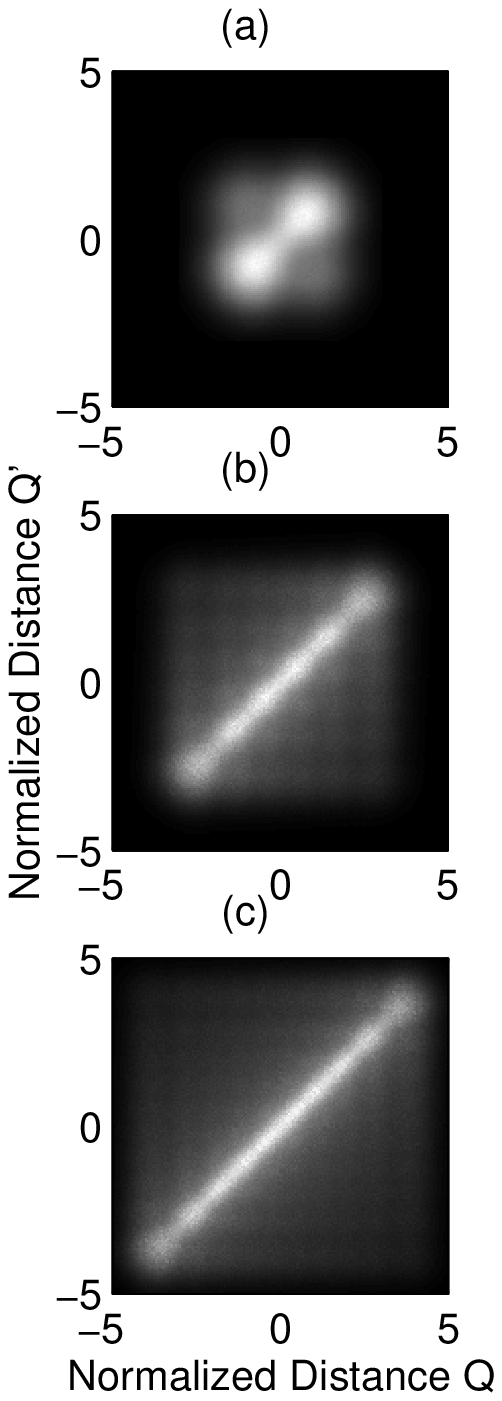}
\begin{center}
{\bf Figure 2}
\end{center}
\newpage
\includegraphics*[width=1.0\columnwidth]{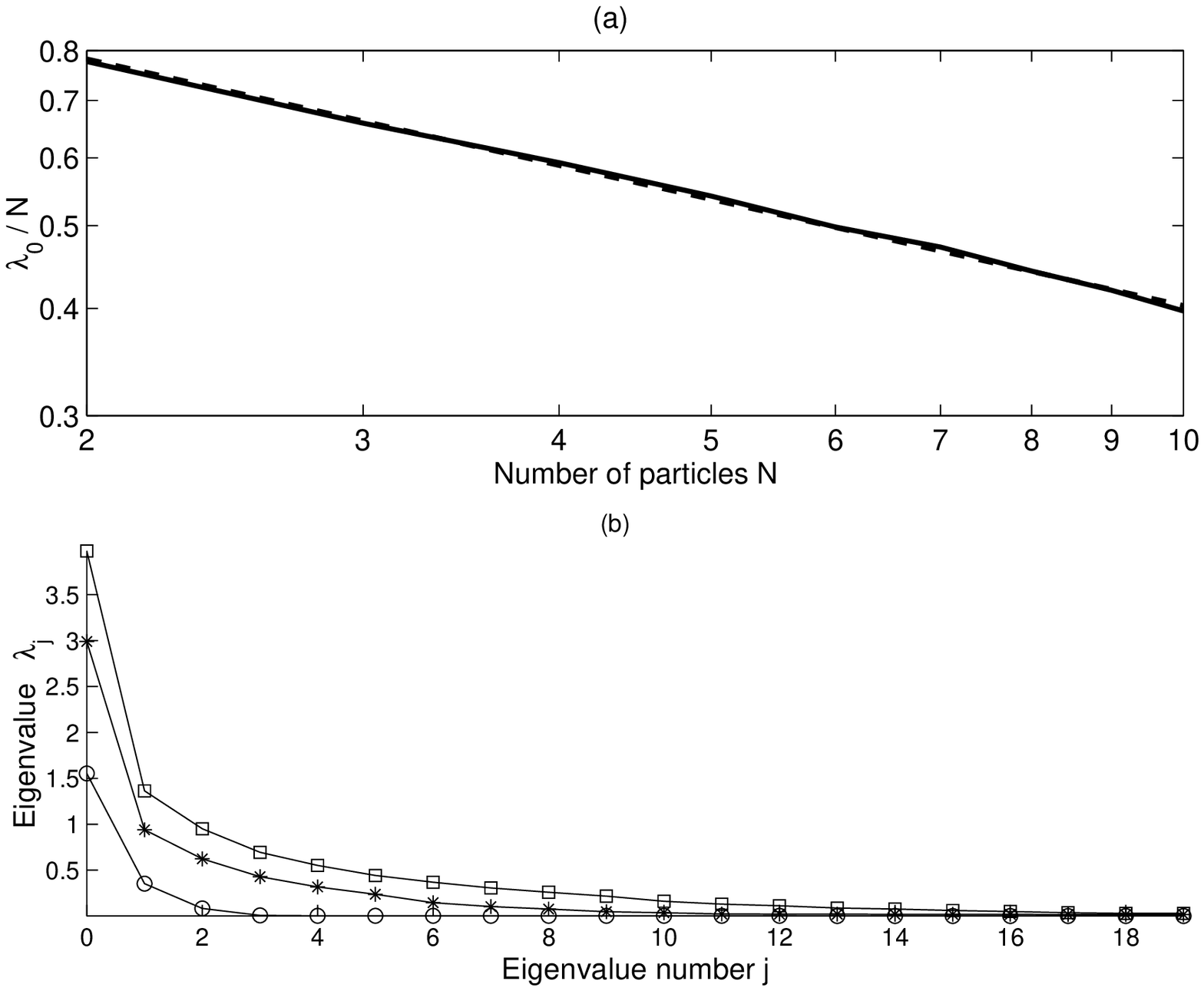}
\begin{center}
{\bf Figure 3}
\end{center}
\newpage
\includegraphics*[width=1.0\columnwidth]{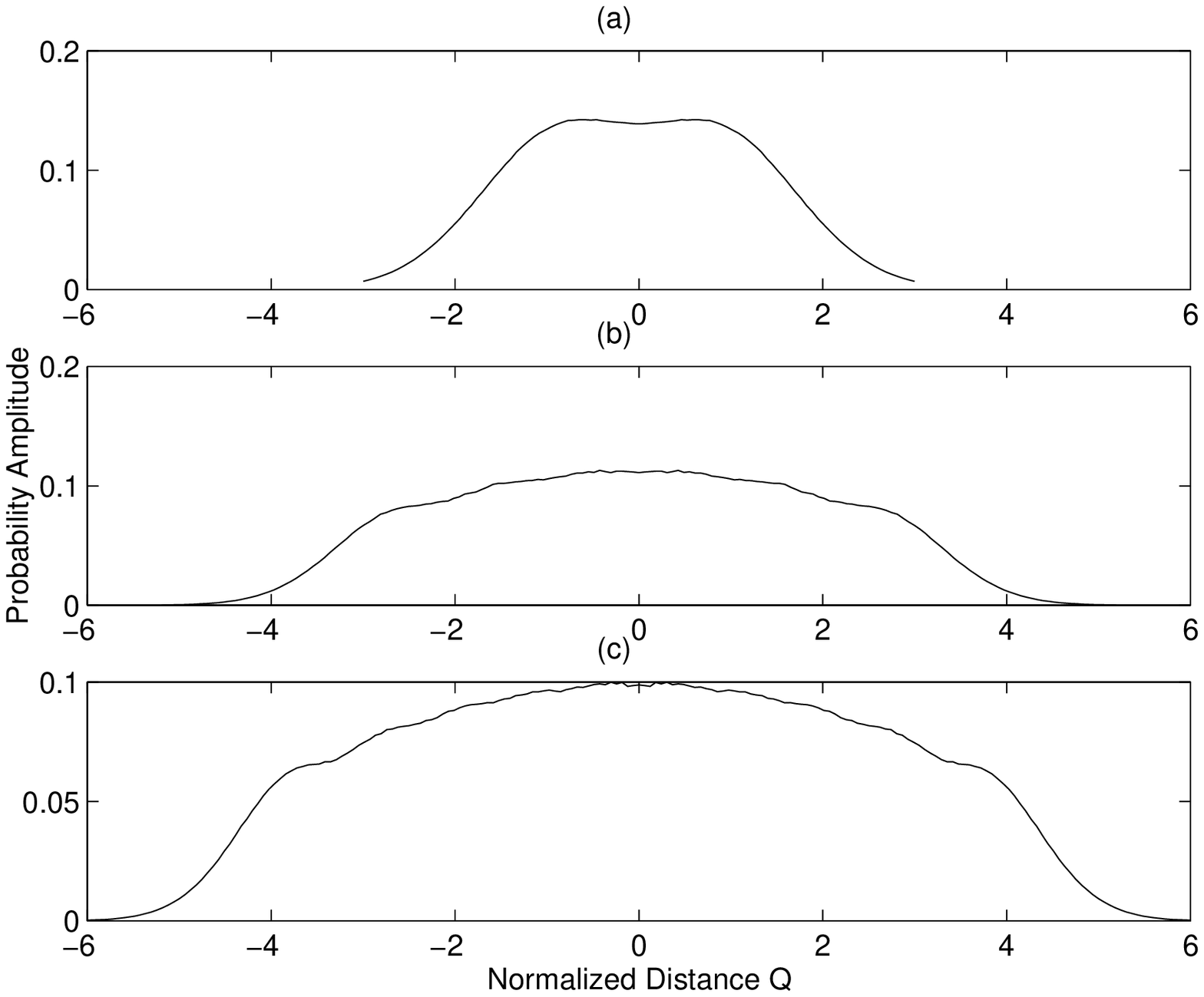}
\begin{center}
{\bf Figure 4}
\end{center}
\newpage
\includegraphics*[width=1.0\columnwidth]{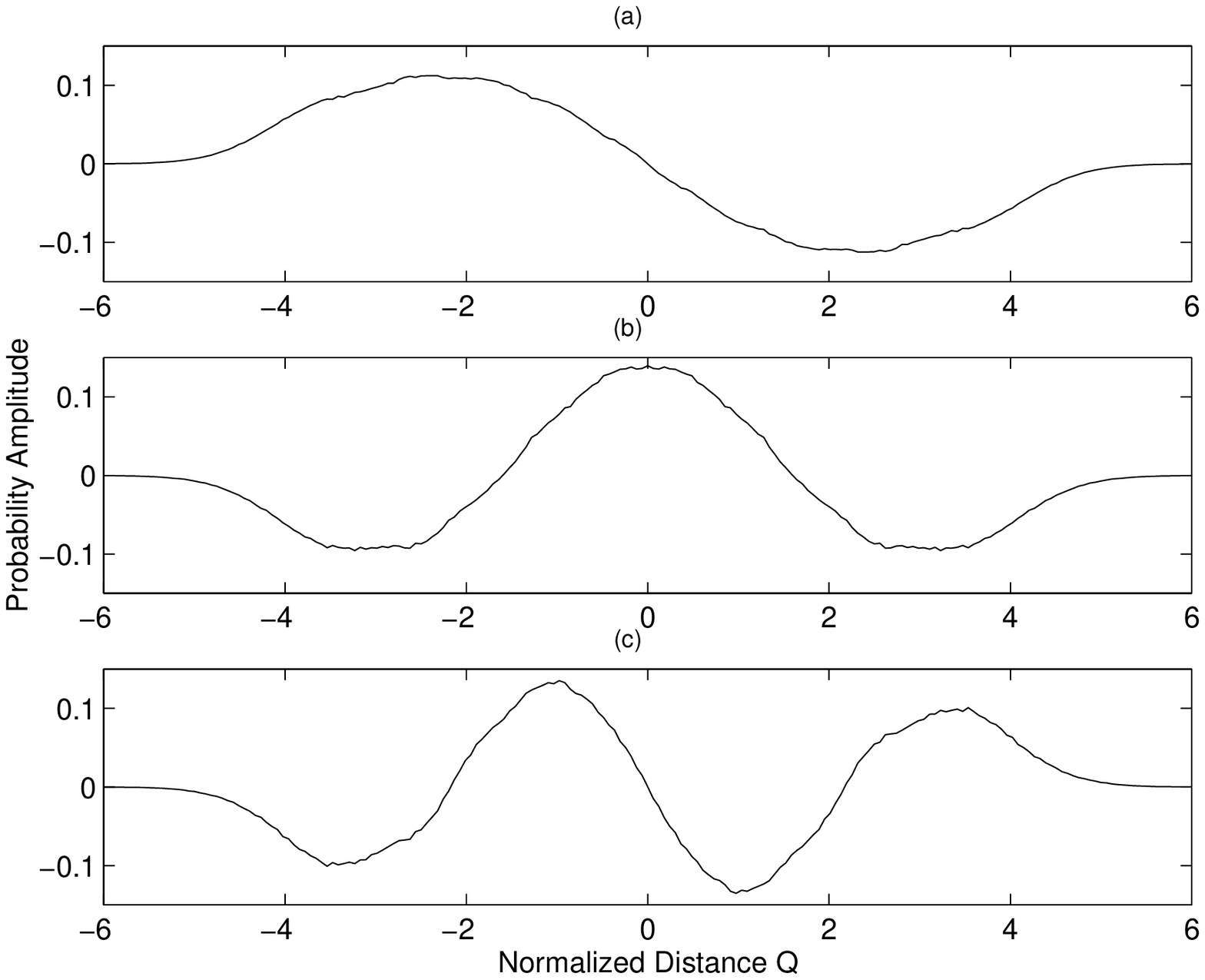}
\begin{center}
{\bf Figure 5}
\end{center}
\newpage
\includegraphics*[width=1.0\columnwidth]{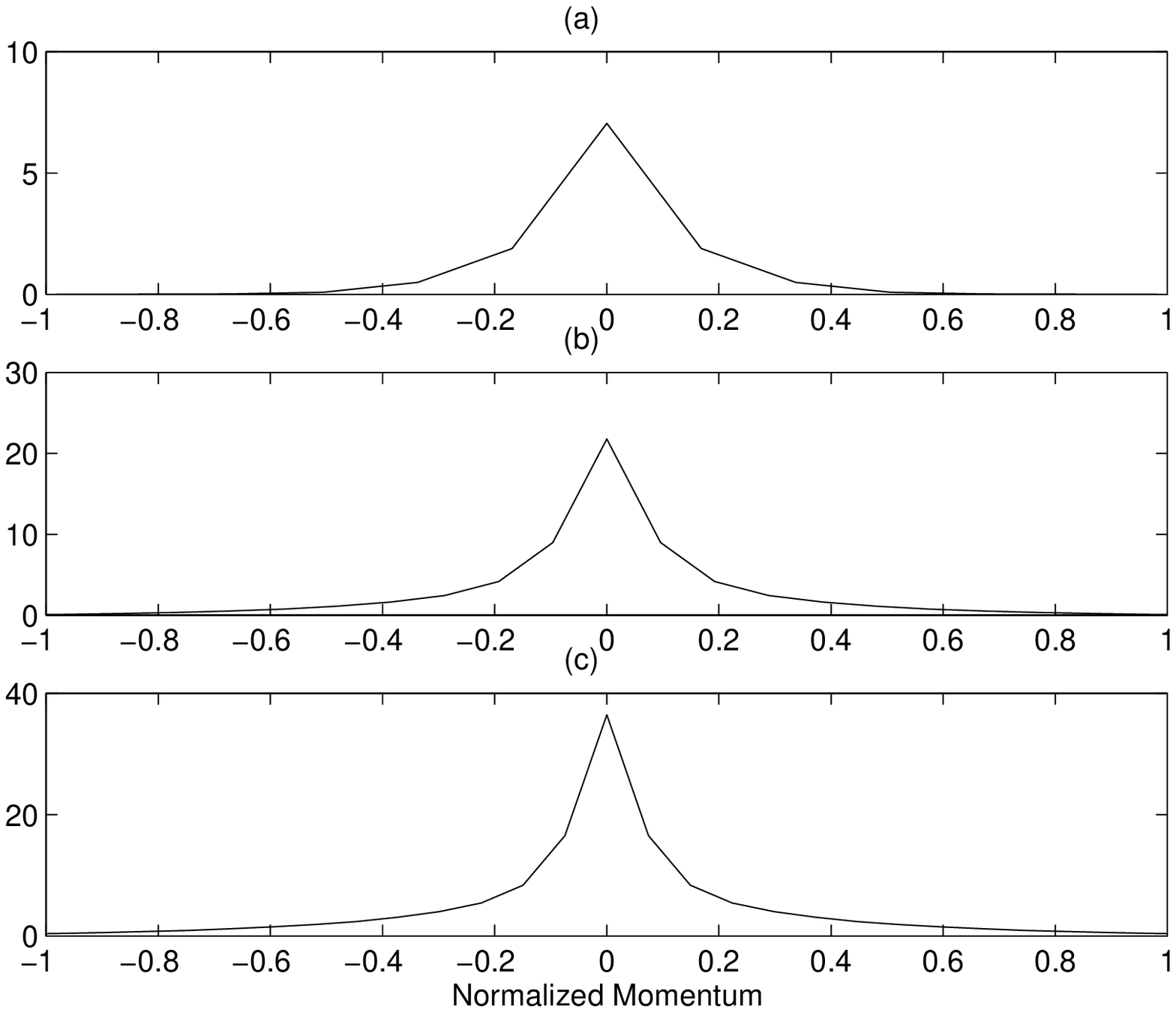}
\begin{center}
{\bf Figure 6}
\end{center}
\end{document}